\documentclass[3p,times]{elsarticle}
\usepackage{ecrc}
\usepackage{epstopdf}
\volume{00}
\firstpage{1}
\journalname{Nuclear Physics A}

\runauth{S.A. Voloshin and R. Belmont}

\jid{nupha}
\jnltitlelogo{Nuclear Physics A}

\usepackage{graphicx}
\usepackage{amsmath}
\usepackage{MnSymbol}
\usepackage{float}

\usepackage{lineno}

\def\snn{\mbox{$\sqrt{s_{_{\rm NN}}}$}}

\def \deta {\Delta \eta} 
\def \dphi {\Delta \phi} 
 
\newcommand{ \be }{\begin{eqnarray}}
\newcommand{ \ee }{\end{eqnarray}}
\newcommand{ \la }{\langle}
\newcommand{ \ra }{\rangle}

\newcommand{ \mean }[1]{\left\langle #1 \right\rangle}

\newcommand{\pt}{p_T}
\newcommand{\corr}{\llangle \cos[n(\phi_1-\Psi_n)]\,c_3 \rrangle }

\begin{document}

\begin{frontmatter}



\title{Measuring and interpreting charge dependent anisotropic flow}

\author{Sergei A. Voloshin\corref{cor1}}
\author{Ronald Belmont}
\address{Department of Physics and Astronomy, Wayne State University,
666 W. Hancock,   Detroit, Michigan 48201}

\cortext[cor1]{Corresponding author. E-mail: sergei.voloshin@wayne.edu}




\begin{abstract}
The Chiral Magnetic Wave (CMW)~\cite{Burnier:2011bf} predicts a
dependence of the positive and negative particle elliptic flow on the
event charge asymmetry. Such a dependence has been observed by the
STAR Collaboration~\cite{Wang:2012qs}. 
However, it is rather difficult to interpret the results of
this measurement, as well as to perform cross-experiment comparisons, due to the
dependence of the observable on experimental inefficiencies and the kinematic
acceptance used to determine the net asymmetry.
We propose another observable that is
free from these deficiencies. It also provides possibilities for
differential measurements clarifying the interpretation of the
results. We use this new observable to study the effect of the local
charge conservation that can mimic the effect of the CMW in charge
dependent flow measurements.

\end{abstract}


\end{frontmatter}



\section{Introduction}
\label{intro}
The STAR observation~\cite{Wang:2012qs} of the dependence of the
elliptic flow of positively and negatively charged particles, $v_2^+$
and $v_2^-$ on the event charge asymmetry $A_\pm=(N_+-N_-)/(N_++N_-)$
(measured in STAR TPC acceptance) has attracted strong attention as it
was found to agree, at least qualitatively, with the predictions for
the Chiral Magnetic Wave (CMW)~\cite{Burnier:2011bf}. Note that more
recent theoretical calculations of the CMW effect are rather
controversial -- they range from a very modest signal that would be
difficult to detect~\cite{Hongo:2013cqa,Taghavi:2013ena} to a ``full
strength'' effect~\cite{Yee:2013cya} explaining the entire signal
measured by STAR.  Naively, it is very difficult to imagine any
background mechanisms that would lead to the dependence observed by
STAR and there exist a very limited number of possible explanations
proposed so far.  Perhaps the most interesting of these explanations
is the result of Bzdak and Bozek~\cite{Bzdak:2013yla} who tried to
incorporate the effect of the local charge conservation (LCC) in the
hydrodynamic calculations.  Their calculations show that LCC can
indeed be responsible for at least a large part of the observed
effect. Unfortunately it is difficult to trace from the final result
the real mechanism -- how and why LCC actually contributes to this
observable.

Note that the importance of the charge dependent anisotropic flow
study goes far beyond the CMW search. In heavy ion collisions the
strength of the magnetic and electric fields can reach extremely high
values $eB \sim eE \sim m_\pi^2$. The lifetime of the magnetic field
is not exactly known and depends on the electric conductivity of the
system. One of the recent calculations~\cite{Gursoy:2014aka} shows
that the lifetime of the magnetic field can be quite large, resulting
in significant modification of charged particle momentum
distributions. In particular, the deflection of the charged particle
trajectories in the reaction plane can contribute to the elliptic flow
development. Speculating that the effect happens at the quark level it
would lead to $v_{2,u}>v_{2,d} \ge v_{2,s}$. Then one should expect
that $v_{2,p}>v_{2,n} \ge v_{2.\Lambda}$, $v_{2,\pi^+}=v_{2,\pi^-} =
v_{2,\pi^0}$, $v_{2,K^+}=v_{2,K^-} > v_{2,K^0}$. Interestingly,
the recent ALICE measurements of charged and neutral Kaons
flow~\cite{Abelev:2014pua} would be consistent with such
relations. The precise measurements of such relations would allow us to
say much more about the role of the magnetic field in the system
evolution. In a very interesting scenario proposed by
S.~Pratt~\cite{Pratt:2012dz,Pratt:2013xca} the quark production
happens in two ``waves'' with different ratios of the strange and light
quarks produced at the very early times and at the hadronization
stage. It would be extremely interesting to see the difference in
electro-magnetic field effect on quarks produced at different times.

One of the difficulties in comparing the flow results as a function of
the event charge asymmetry $A_\pm$ obtained in different
systems and/or at different collision energies is the dependence of
$A_\pm$ on the detector acceptance as well
as on experimental tracking efficiencies. The new technique discussed
below is free from both of these problems, as the corresponding
observable is efficiency independent and represents the dependence on
the ``charge asymmetry'' in a differential way, allowing the comparison
to be performed exactly in the (overlap) regions covered by different
experiments. The differential character of the observable allows also
more detailed study of the underlying physics mechanisms. We
demonstrate this below using the new observable to study LCC effects
on the charge dependent anisotropic flow measurements.

\section{Correlator $\corr$} 

The meaning of this correlator is the anisotropic flow of a particle
as a function of the mean charge of another particle in a given
momentum window. Usually we study the dependence on the pseudorapidity
difference between the two particles. The notations here are the
following: the particle for which the flow is measured is called
particle ``1'', the event plane is estimated with particle ``2'', and
the average charge is calculated for the particle ``3''. Note that the
average charge of particle ``3'' over the acceptance is $\la
c_3\ra=A_\pm$. Evaluation of the anisotropic flow is assumed to follow
the standard procedure (e.g. evaluation of the event plane and
subsequent correction for the reaction plane resolution, or the use of
the scalar product method, etc.) and is not discussed here. The
correlator is a 3-particle (point) correlator, and double brackets
mean the cumulant. It assumes subtraction of contributions from all
``reducible'' correlations. One has to take into account here a
non-trivial character of the correlations of particle ``1'' - it is
correlated with the flow angle $\Psi_n$ via azimuth and with particle
``3'' via both charge and momentum.
\be \corr = \la \cos[n(\phi_1-\Psi_n)] \, c_3 \ra - \la
\cos[n(\phi_1-\Psi_n)] \ra \, \la c_3 \ra_1 \ee 
where single brackets
means average over particles and events. $\la c_3 \ra_1$, a function
of $\deta=\eta_1-\eta_3$, is an average of the charge of particle ``3''
under condition of particle ``1'' (of a given charge) to be observed.
This function is proportional to the charge balance function
$B(\Delta\eta)$.

\begin{figure}[ht]
\begin{center}
\includegraphics*[width=0.4\textwidth]{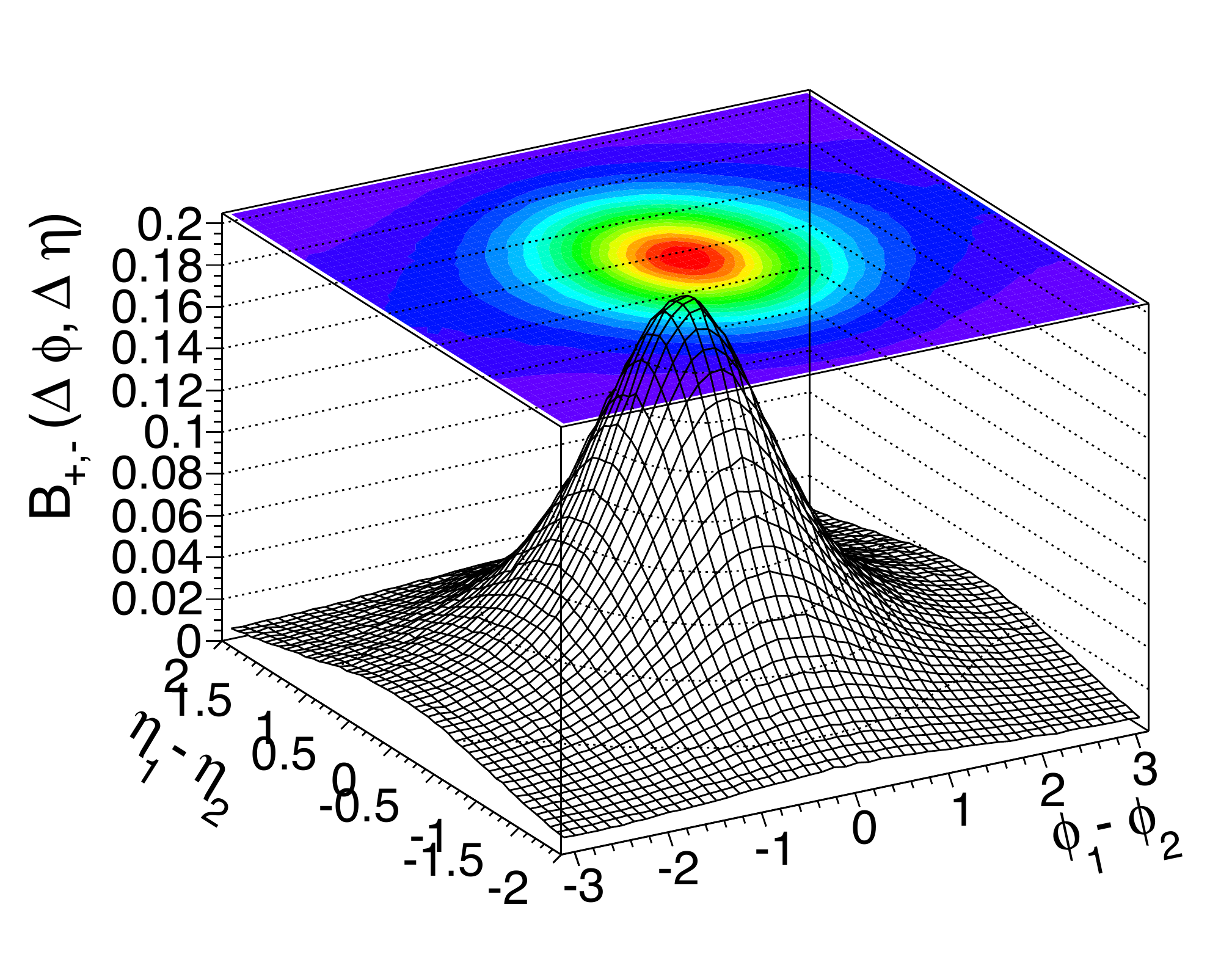}
\hspace{0.1\textwidth}
\includegraphics*[width=0.4\textwidth]{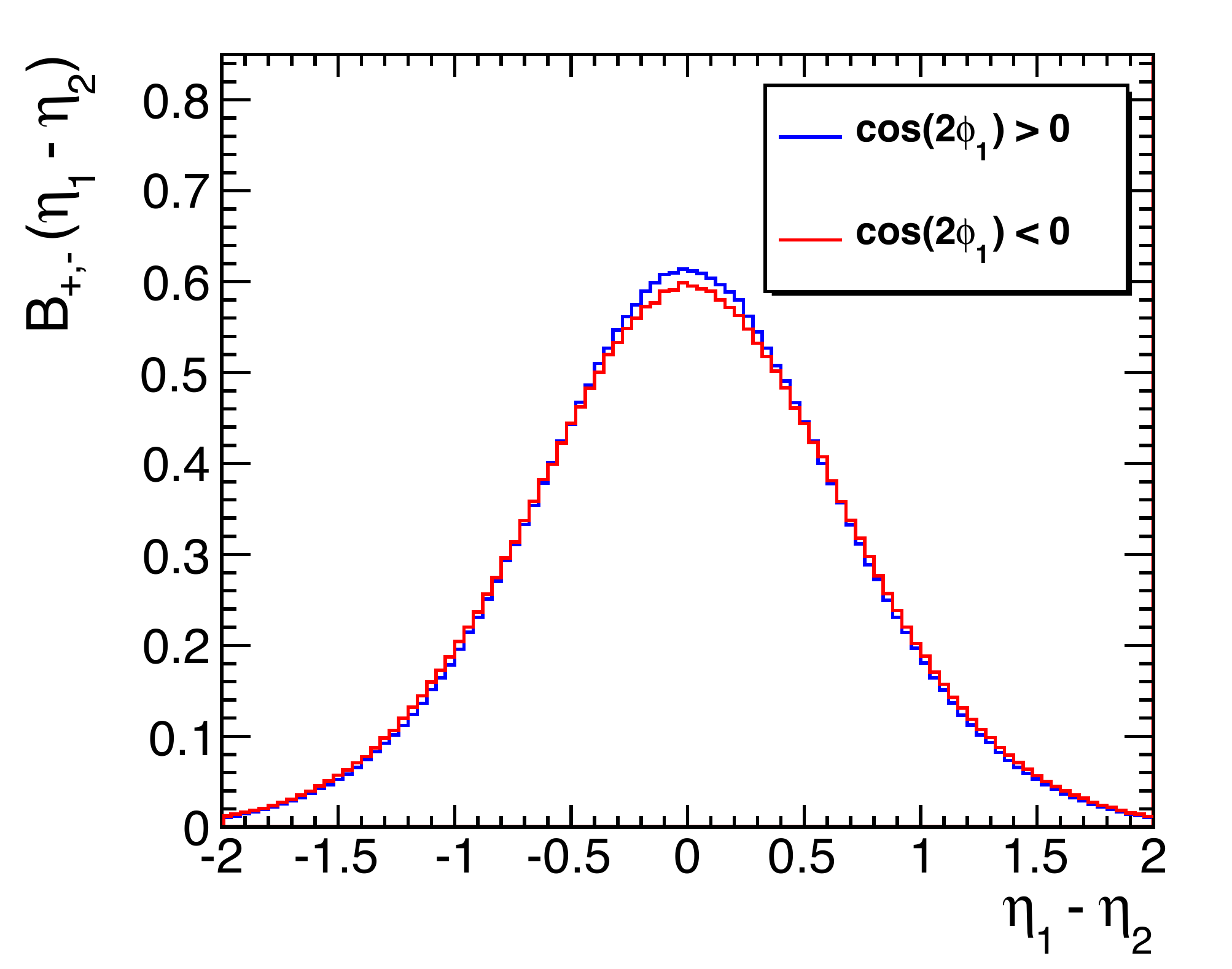}
\caption{
Left: Charge balance function as a function of $\deta,\;\dphi$. For
a description of the model see the text. Right: $\deta$ projections of the
balance functions for the particle ``1'' emitted in-plane and out-of-plane.}
\label{fig:bf}
\end{center}
\end{figure}

\section{The Blast Wave model and effect of the local charge conservation}

To demonstrate the basic features of the proposed correlator as well
as to get estimates of the LCC effect, we employ a simple Blast Wave
model. The model assumes the source at constant temperature $T$, and
with the radial boost rapidity and the number of elementary cells,
$n_c$, modulated according to 
\be \rho_t(r,\phi_b)=\rho_{t,max}\frac{r}{r_{max}(\phi_b)} \left(
1+\rho_2 \cos[2(\phi_b-\Psi_2)] +\rho_3\cos[3(\phi_b-\Psi_3)]\right),
\\ \frac{dn_c}{d\phi_b}\propto
1+s_2\cos[2(\phi_b-\Psi_2)]+s_3\cos[3(\phi_b-\Psi_3)], 
\ee 
where $\phi_b$ is the boost angle.  We assume all particles to have
pion mass, and the flow angles $\Psi_2$ and $\Psi_3$ are assumed to be
uncorrelated.  We use $T=100$~MeV, $\rho_{t,max}=1.1$, $s_2=0.1$,
$s_3-0.06$, $\rho_2=0.06$ and $\rho_3=0.03$.  The parameters of the
model have not been fit to the data on anisotropic flow, but selected
such that it (very) roughly correspond to Pb--Pb collisions at 
$\snn=2.76$~TeV and 30\% centrality.  The charged particle density,
also needed for correlator calculations (to account for statistical
dilution of the correlations), chosen to be $dN_{ch}/d\eta=420$.

\begin{figure}
\begin{center}
\includegraphics*[width=0.4\textwidth]{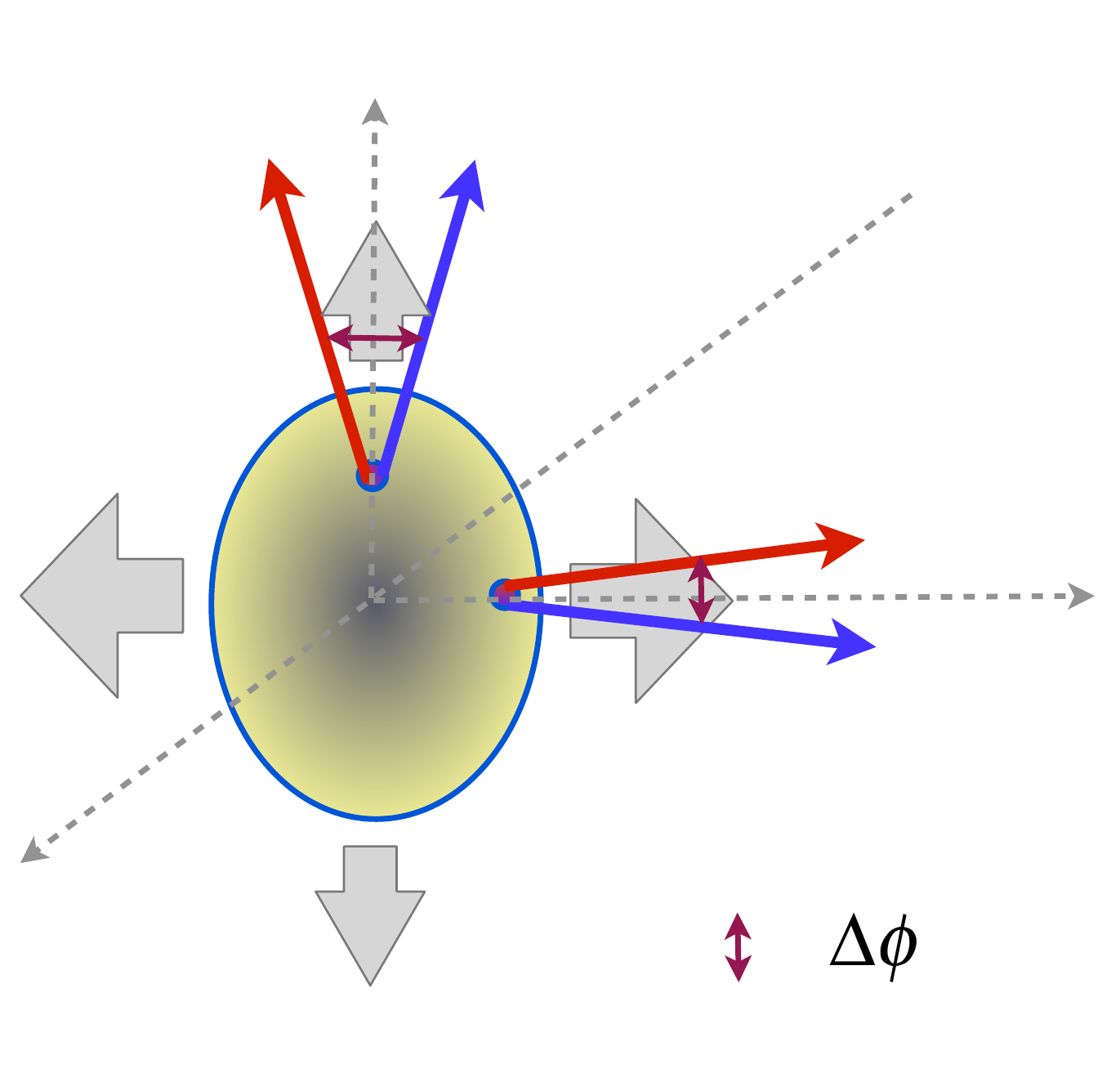}
\hspace{0.1\textwidth}
\includegraphics*[width=0.4\textwidth]{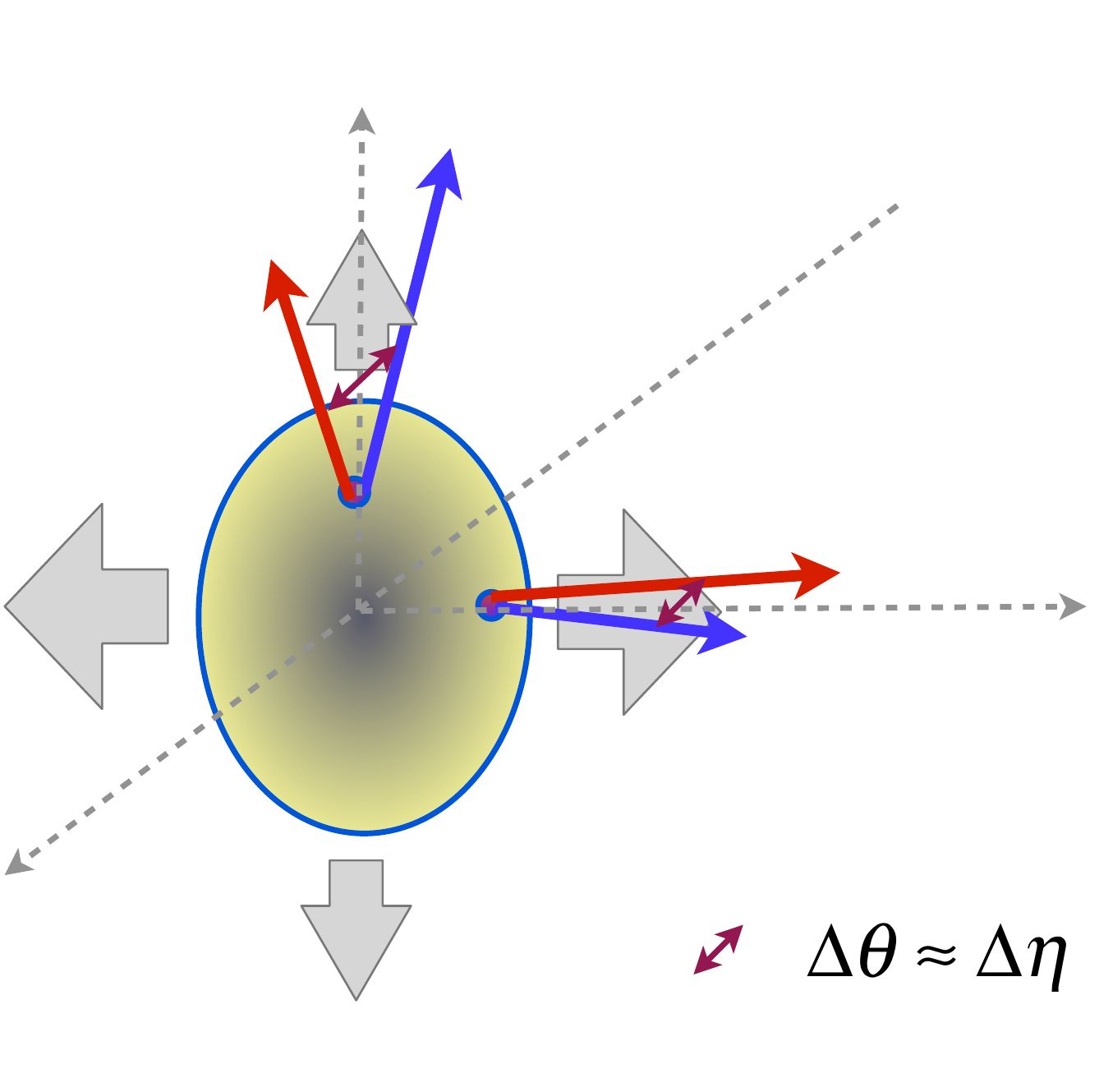}
\caption{Effect of the radial boost on two particle angular difference in
  azimuth (left panel) and pseudorapidity (right panel).
}
\label{fig:mechanism}
\end{center}
\end{figure}

To study the LCC effect we assume that both balancing charges are
produced at the same spatial point and become correlated due to the
radial (azimuthally modulated) boost according to the initial radial
position. We assume that the particle production is boost invariant
in the longitudinal direction, and the particle separation in
(pseudo)rapidity is due to their thermal velocities, collimated 
due to the radial boost~\cite{Voloshin:2003ud}.  The pair distribution
(of the balancing charge relative to the position of the
first particle) as a function of separation in pseudorapidity and
azimuthal angle -- the balance function -- is shown in
Fig.~\ref{fig:bf}a. Figure~\ref{fig:bf}b shows two projection of the
balance function on pseudorapidity difference for pairs selected in
accordance to the emission angle of the first particle -- either
preferentially in-plane or out-of-plane. The widths of the two
distribution are different reflecting the effect of elliptic flow --
stronger radial expansion in-plane; see also Fig.~\ref{fig:mechanism}.
This effect is very similar to the contribution of LCC to the
3-particle correlator~\cite{Voloshin:2004vk,Abelev:2009ac} used in
search for the Chiral Magnetic Effect~\cite{Kharzeev:2007jp}, and
first discussed in~\cite{Schlichting:2010qia} -- stronger focusing of
particle pairs emitted in-plane compared to those emitted
out-of-plane.

\begin{figure}
\begin{center}
\includegraphics*[width=0.4\textwidth]{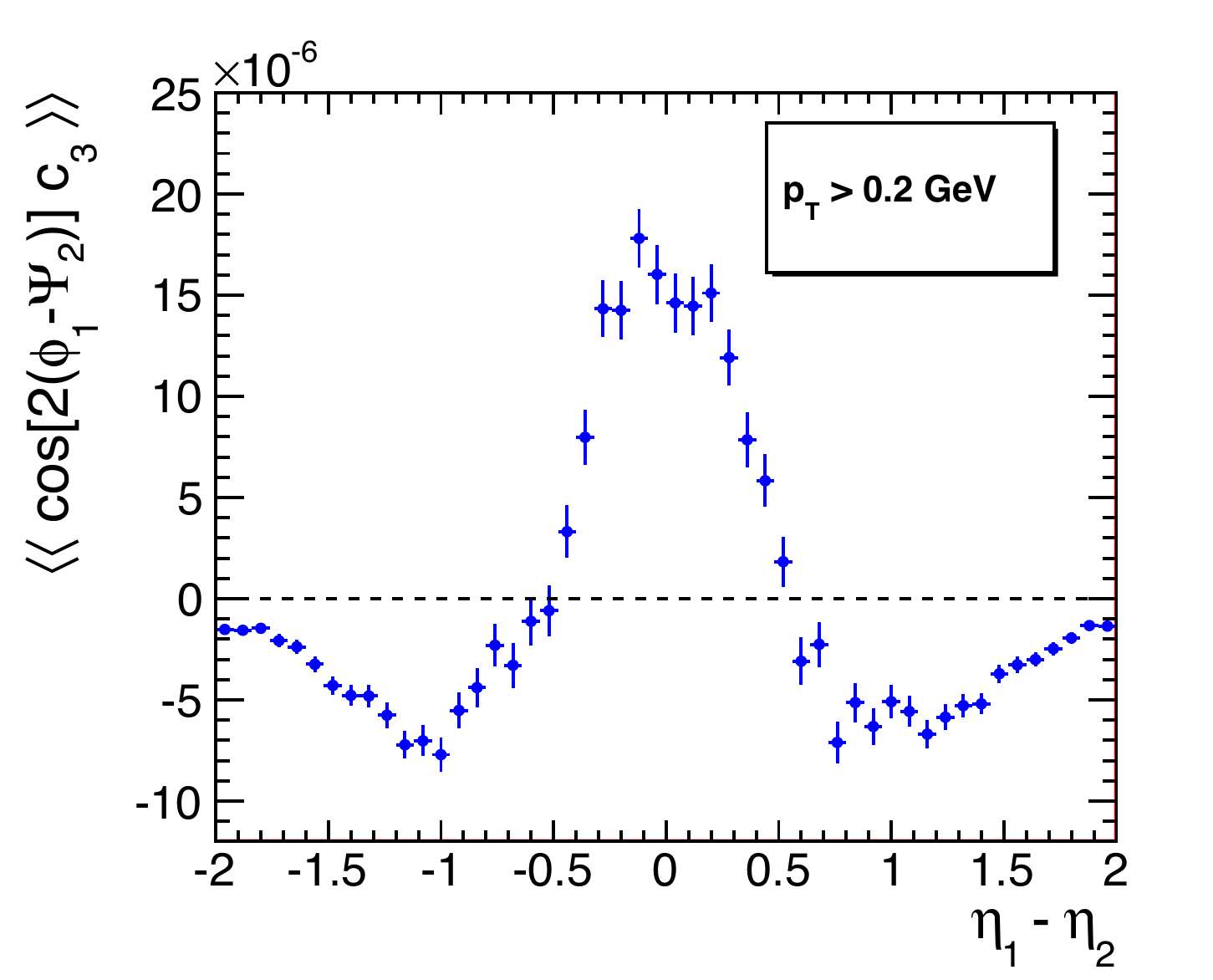}
\hspace{0.1\textwidth}
\includegraphics*[width=0.4\textwidth]{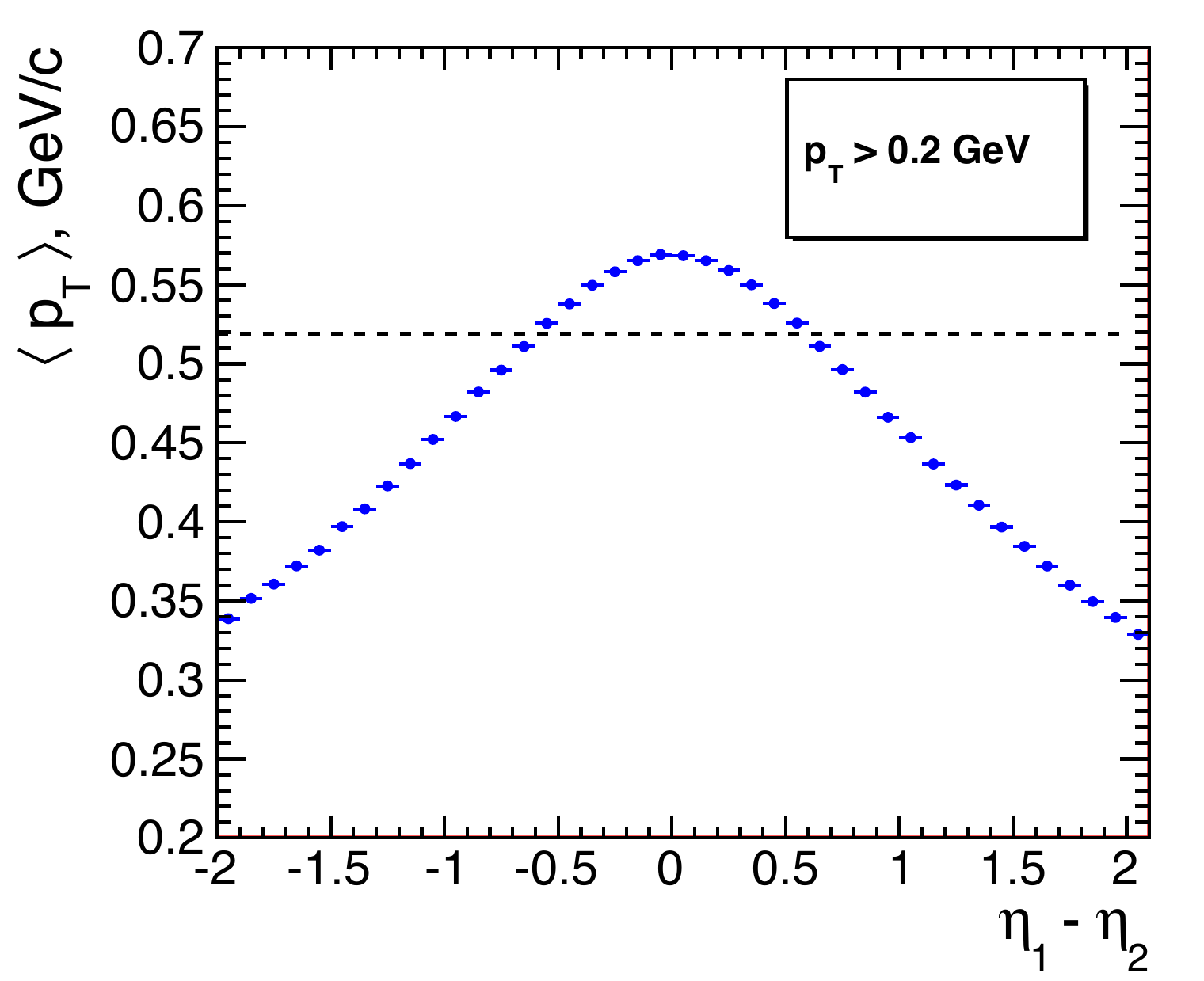}
\vspace*{-1mm}
\caption{
Left: The correlator $\corr$ as a function of $\Delta\eta$. 
Right: Mean $\pt$ of the balancing charge particle.}
\label{fig:corr}
\end{center}
\end{figure}

Figure~\ref{fig:corr}a presents one of the main results of this
presentation -- the correlator as function of the pair (particles
``1'' and ``3'') separation in pseudorapidity. This results is an
interplay of the two effects, a stronger correlation of
balancing charges in-plane compared to out-of-plane, and the
statistical ``dilution'' of the correlation due to uncorrelated
background. Note that the correlator becomes negative at $\Delta\eta
\gtrsim 0.6$ -- 0.7.  It then approaches zero due to too few balancing
particles at large rapidity separation.
Figure~\ref{fig:corr}b shows the mean $\pt$ of the balancing charge as
a function of $\Delta\eta$. The dashed line indicates the inclusive
value of $\mean{\pt}$. Higher than average $\pt$ values are observed
at small rapidity separation, corresponding to stronger radial
flow. One can use this effect as an additional test of the effect of
the LCC, e.g. by measuring $\mean{p_{T,3} \, c_3} -
\mean{p_{T,3}}\mean{c_3}_1$ as a function of $\eta_1-\eta_3$.

The LCC effect should be also seen in higher harmonic correlators,
while the effects of CME and CMW should be
minimal~\cite{Voloshin:2011mx}.  According to our Blast Wave
calculations the third harmonic correlator is very similar in shape to
that shown in Fig.~\ref{fig:corr} with about thee times smaller an
amplitude compared to the second harmonic.

\section{Conclusions}

In summary, we propose a new correlator, $\corr$, to study the charge
dependence of the elliptic flow. This correlator is directly sensitive
to the CMW, but being efficiency independent and differential in
nature, allows more thorough investigation of the underlying physical
mechanisms. Our study of the LCC effects in the Blast Wave model
reveals that stronger in-plane boost due to elliptic flow leads to
narrower distribution in $\Delta\eta$ of the balancing charges, which
propagates directly to the measured correlation. The results are in
rough qualitative agreement with ALICE Collaboration
measurements~\cite{BelmontQM2014}, but more detail calculations, as
well as measurements, are needed to draw more definite conclusion
about the LCC role in a measured signal.  Such Blast Wave calculations
can be also used to directly relate the contribution of the LCC to the
observable used for CMW search as well as the ones for the CME search,
which would be an important cross check.

This material is based upon work supported by the U.S. Department of 
Energy Office of Science, Office of Nuclear Physics under Award 
Number DE-FG02-92ER-40713.


\begin{thebibliography}{00}

\bibitem{Burnier:2011bf} 
  Y.~Burnier, D.~E.~Kharzeev, J.~Liao and H.~-U.~Yee,
  Phys.\ Rev.\ Lett.\  {\bf 107}, 052303 (2011)
  [arXiv:1103.1307 [hep-ph]].

\bibitem{Wang:2012qs} 
  G.~Wang [STAR Collaboration],
  Nucl.\ Phys.\ A {\bf 904-905}, 248c (2013)
  [arXiv:1210.5498 [nucl-ex]].

\bibitem{Hongo:2013cqa} 
  M.~Hongo, Y.~Hirono and T.~Hirano,
  arXiv:1309.2823 [nucl-th].

\bibitem{Taghavi:2013ena} 
  S.~F.~Taghavi and U.~A.~Wiedemann,
  arXiv:1310.0193 [hep-ph].

\bibitem{Yee:2013cya} 
  H.~-U.~Yee and Y.~Yin,
  Phys.\ Rev.\ C {\bf 89}, 044909 (2014)
  [arXiv:1311.2574 [nucl-th]].

\bibitem{Bzdak:2013yla} 
  A.~Bzdak and P.~Bozek,
  Phys.\ Lett.\ B {\bf 726}, 239 (2013)
  [arXiv:1303.1138 [nucl-th]].


\bibitem{Gursoy:2014aka} 
  U.~Gursoy, D.~Kharzeev and K.~Rajagopal,
  Phys.\ Rev.\ C {\bf 89}, 054905 (2014)
  [arXiv:1401.3805 [hep-ph]].

\bibitem{Abelev:2014pua} 
  B.~B.~Abelev {\it et al.}  [ALICE Collaboration],
  arXiv:1405.4632 [nucl-ex].

\bibitem{Pratt:2012dz} 
  S.~Pratt,
  Phys.\ Rev.\ Lett.\  {\bf 108}, 212301 (2012)
  [arXiv:1203.4578 [nucl-th]].

\bibitem{Pratt:2013xca} 
  S.~Pratt,
  PoS CPOD {\bf 2013}, 023 (2013)
  [arXiv:1304.2442 [nucl-th]].

\bibitem{Voloshin:2003ud} 
  S.~A.~Voloshin,
  Phys.\ Lett.\ B {\bf 632}, 490 (2006)
  [nucl-th/0312065].

\bibitem{Voloshin:2004vk} 
  S.~A.~Voloshin,
  Phys.\ Rev.\ C {\bf 70}, 057901 (2004)
  [hep-ph/0406311].

\bibitem{Abelev:2009ac} 
  B.~I.~Abelev {\it et al.}  [STAR Collaboration],
  Phys.\ Rev.\ Lett.\  {\bf 103}, 251601 (2009)
  [arXiv:0909.1739 [nucl-ex]].

\bibitem{Kharzeev:2007jp} 
  D.~E.~Kharzeev, L.~D.~McLerran and H.~J.~Warringa,
  Nucl.\ Phys.\ A {\bf 803}, 227 (2008)
  [arXiv:0711.0950 [hep-ph]].

\bibitem{Schlichting:2010qia} 
  S.~Schlichting and S.~Pratt,
  Phys.\ Rev.\ C {\bf 83}, 014913 (2011)
  [arXiv:1009.4283 [nucl-th]].

\bibitem{Voloshin:2011mx} 
  S.~A.~Voloshin,
  Prog.\ Part.\ Nucl.\ Phys.\  {\bf 67}, 541 (2012)
  [arXiv:1111.7241 [nucl-ex]].

\bibitem{BelmontQM2014}
 R.~Belmont for the ALICE Collaboration, these proceedings.









\end{thebibliography}
\end{document}